\documentclass[12pt]{iopart}

\usepackage{graphicx}
\begin{document}

\title[]{Reduction in energy dissipation rate with increased effective applied field}

\author{Zden\v{e}k Jan\accent23 u and Franti\v{s}ek Soukup}


\address{Institute of Physics of The CAS, Na Slovance 2, CZ-182 21, Prague, Czech Republic} \ead{janu@fzu.cz}

\begin{abstract}
Dynamics of a response of type-II superconductors to a time-varying magnetic field can exhibit rate independent or rate dependent hysteresis. An energy dissipation rate in a superconductor placed in a time varying magnetic field depends on its waveform and type of hysteresis which depends on temperature. The same waveform may reduce the energy dissipation rate in the case of true hysteresis while may increase the energy dissipation rate in the case of dynamic hysteresis compared with an energy dissipation rate in a pure sinusoidal field. We present experimental data which confirm the energy dissipation rate calculated using the critical state theory for the case of rate independent hysteresis and limiting behavior in normal state for the case of rate dependent hysteresis.
\end{abstract}


\maketitle

\section{Introduction}
Magnetic hysteresis in type-II superconductors in an external time varying magnetic field may vary between the rate independent hysteresis and the rate dependent hysteresis depending on temperature \cite{Campbell}. The rate dependent hysteresis and the rate independent hysteresis bring about a markedly different energy dissipation rate depending on a waveform of the external field. When, e.g., a field of the third harmonic frequency in phase with the pure sinusoidal field is added, a peak value of the field is trimmed, the field rate changes, and the effective field increases. We compare the dissipation rates in the case of pure sinusoidal and first plus third harmonic or (approximated) square waves. One can reduce the ac losses when the superconductor is in the critical state by adding harmonics, even if the effective applied field is increased. On the contrary, we observe an increase in dissipation with the number of harmonics when the superconductor is in the normal state, as expected.

A growing application potential of thin superconducting films employed, for example, in recent second-generation high temperature superconducting wires or superconducting electronics, makes an exploration of the field waveform dependent energy dissipation rate in the thin films relevant. Recently, the critical-state models for thin circular disks and strips in a transverse magnetic field were developed \cite{Brandt93,Clem94}. A validity of these models was proved later \cite{Janu09,Janu14}. The models give complete analytical expressions needed to calculate magnetization loops in a time varying magnetic field. On the basis of these magnetization loops the energy dissipation rate may be calculated for the case of the rate independent hysteresis which occurs at temperatures below the critical depinning temperature. At temperatures above the critical depairing temperature, a superconductor is in a normal state with ohmic conductivity. The magnetization loops manifest the rate dependent hysteresis. For a low frequency ac field the limiting behavior of the energy dissipation rate may be obtained on the basis of the eddy current model \cite{Khoder91}. In a temperature interval between the critical depinning temperature and critical depairing temperature a hysteresis is a rate dependent. A shape of the magnetization loops depends both on a peak value and a rate of the field because of screening current damping of which time dependence changes from logarithmical to exponential with increasing temperature \cite{Brandt97}.

Our experimental results support the foregoing hypothesis. We have measured temperature dependence of the magnetic moment of a thin Nb film subjected to an external time varying magnetic field which waveform was synthesized to modify a peak value and rate. The energy dissipation rate calculated on the basis of these measurements is in agreement with the predicted behavior.

\section{Theory and experiment}

The Nb film of thickness 250 nm was sputtered using a dc magnetron on 400 nm thick SiO$_2$ layer thermally grown on a Si single crystal wafer \cite{May99}. The samples are squares with dimensions 5 $\times$ 5 mm$^{2}$ cut of the wafer. The sample is mounted on a sapphire sample holder temperature of which is controlled electronically. The component $m$ of the induced magnetic moment parallel to the field $H$ oriented perpendicularly to the film is measured using a continuous reading SQUID magnetometer \cite{Janu09,Janu14}.

It is convenient to Fourier analyze signals periodic in a time domain. We apply the discrete fast Fourier transform to calculate spectra $\mathcal{M}(f)$ and $\mathcal{H}(f)$ of digitized signals $m(t)$ and $H(t)$. The frequency spectrum of the applied pure sinusoidal field $H(t)=H_{1}\cos(2\pi f_{1} t)$ has the complex amplitude $\mathcal{H}(f_{1}) = \mathcal{H}'(f_{1}) + i\mathcal{H}''(f_{1}) = H_{1} + i0$. The nonlinear $m(H)$ dependence produces harmonics in the magnetic moment. We denote the $n$th harmonic component $\mathcal{M}_{n}(f_{1}) \equiv \mathcal{M}(nf_{1})$. The mean energy stored in the sample is the real part of $-(\mu_{0}/2)\mathcal{H}^*(f_{1})\mathcal{M}_1(f_{1})$ and energy dissipated in the sample per ac field cycle is the imaginary part of $-(\mu_{0}/2)\mathcal{H}^*(f_{1})\mathcal{M}_1(f_{1})$. The asterisk denotes a complex conjugate amplitude. The mean value of the energy dissipation rate in the sample is $P = f_{1}W(f_{1})$, where $W(f_{1}) = \pi \mu_{0} \mathrm{Im}[\mathcal{H}^*(f_{1})\mathcal{M}_{1}(f_{1})]$ is the area of the magnetization hysteresis loop $m(H)$.

Magnetic properties of superconductors are commonly described via external ac susceptibility $\chi_{n}$ related to a magnetic moment as $\chi_{n}/\chi_{0} = \mathcal{M}_{n}/m_{M}$, where $\chi_{0} = |(1/V)(\mathrm{d}m/\mathrm{d}H)_{H=0}|$ is the initial susceptibility dependent on a shape of the sample and its orientation in a field and $m_{M} = \chi_{0} V H$ is the magnitude of the magnetic moment of the sample in the case of perfect screening \cite{Goldfarb91,Chen10}.

At temperature below the critical depinning temperature $T_{c}$, vortices are pinned and flux density profiles created in an applied field are quasistatic with the gradient $|\nabla B| = \mu_{0}J_{c} > 0$, where $J_{c}$ is the critical depinning current density. Due to a large aspect ratio of thin film samples in transverse magnetic field configuration the particular in-plane shape almost does not influence the response to an external magnetic field and normalized magnetization loops for the critical-state differ by less than 1.3\% between disks, strips, and squares over the whole range of applied fields \cite{Brandt97,Chen10}. The only difference is that fitting of data measured on squares to the model for disks gives $J_{c}$ underestimated by 6\% while fitting  to the model for strips gives $J_{c}$ overestimated by 16\%. However, complete analytical expressions know for an initial magnetization curve and magnetization loops of disks and strips allow fast calculations unlike numerical calculations needed for rectangles \cite{Brandt93,Clem94}. The magnetization loops are an analytical function of $H_{p}/H_{d}$, where $H_{p}$ is the peak value of the ac field and $H_{d}$ is the characteristic field. For the disks $H_{d} = J_{c}d/2$ and $\chi_{0} = 8R/3\pi d$, where $R$ is the disk radius and $d$ is the disk thickness \cite{Clem94}.

In the critical state with $J_{c} > 0$, as temperature approaches $T_{c}$, thermal activation of flux lines over a pinning barrier with activation energy $U$ causes the created flux density profiles relax spontaneously by diffusion. The diffusivity $D = (E_{c}/\mu_{0}J_{c})(J/J_{c})^{a-1}$, where $E_{c}/J_{c}$ is a typical resistivity parameter of the material, is a nonlinear function of the screening current $J$. With increasing temperature the creep exponent $a = U / k_{B}T$ falls from $a \rightarrow \infty$ (Bean critical-state) to $a = 1$ (normal state with ohmic conductivity). The critical depinning current density $J_{c}$ drops to zero at $T_{c}$. In the normal state, above the critical depairing temperature $T_{d}$, a superconductor obeys Ohm's law with a conductivity $\sigma$ and diffusivity $D = 1/\mu_{0}\sigma$. Using numerical calculations Brandt has shown that the nonlinear ac susceptibility in a sinusoidal field with an amplitude $H_{p}$ and frequency $f$ may be written as

\begin{equation}
    \label{A}
    \chi \left( f, H_{p} \right) = \mathrm{g} \left( H_{p}/f^{1/(a-1)} + f H_{p}^{1-a}\right),
\end{equation}

\noindent where $\mathrm{g}(f, H_{p})$ is a universal function depending only on geometry \cite{Brandt97}. In Bean critical-state, $\chi$ depends only on $H_{p}$ while, in the normal state, $\chi$ depends only on $f$. As temperature increases from $T_{c}$ to $T_{d}$, nonlinearity in $D(J)$ decreases and $\chi'_{1} / \chi_{0} = \mathcal{M}'_{1}/m_{M}$ is reduced and $\chi''_{1} / \chi_{0} = \mathcal{M}''_{1}/m_{M}$ enhanced at all ac amplitudes $H_{1}$. The maximum of $\mathcal{M}''_{1}/m_{M}$ increases from $\approx 0.24$ for the critical-state to $\approx 0.4$ for the normal state at $R \approx \delta$, where $R$ is the dimension of the sample in a direction perpendicular to the field and $\delta = (\mu_{0} \pi f \sigma)^{-1/2}$ is the skin depth \cite{Landau}. A range of the temperature interval between $T_{c}$ and $T_{d}$ depends on the strength of pinning and an applied field. For weak applied fields these temperatures may coincide.

In the normal state, at temperatures above $T_{d}$, the magnetic moment $\mathcal{M}_{1}$ produced by currents induced in a sample by the changing magnetic field may be calculated on the basis of the eddy current model \cite{Landau}. The magnetization loops are ellipses and for linear $m(H)$ dependence only the fundamental components of the ac magnetic moment are present. In the low frequency limit of an applied ac field, a sample is somewhat transparent for a field when $\delta \gg R$ and components of the ac magnetic moment have the limiting behavior $\mathcal{M}'_{1} \propto - (R/\delta)^{4}$ and $\mathcal{M}''_{1} \propto (R/\delta)^{2} \propto f\sigma$ \cite{Khoder91}. The area of the magnetisation loop $W(f_1) = \pi \mu_{0} \mathcal{M}''_{1}(f_1) H_{1} \propto f_1 \sigma H_{1}^{2}$ increases linearly with a frequency of the external ac field. The mean value of the energy dissipation rate $P = f_1 W(f_1) \propto f_1^{2} \sigma H_{1}^{2}$ grows with the square of frequency.

\begin{figure}
\includegraphics[scale=0.5]{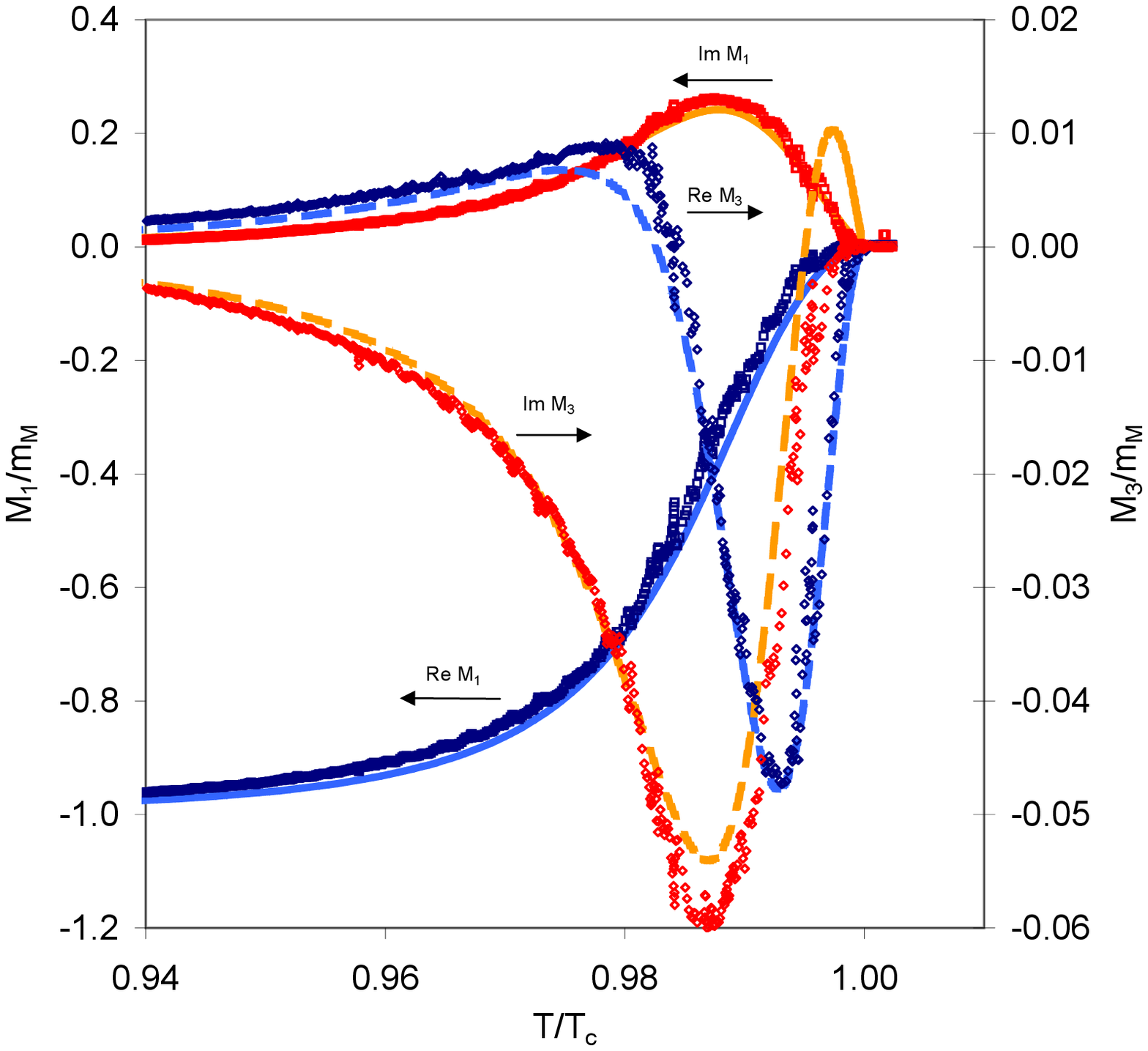}
\caption{\label{FigX1} Temperature dependence of the normalized fundamental and 3rd harmonic ac magnetic moment. Symbols represent experimental data measured in the applied field with the amplitude $\mu_{0} H = 100$ $\mu$T and frequency 1.5625 Hz. The curves represent data calculated on the basis of the model and transformed to temperature domain using effective temperature.}
\end{figure}

Figure \ref{FigX1} shows the experimental ac magnetic moment measured as a function of temperature and theoretical ac magnetic moment of a disk in the critical-state. In order to fit experimental data $[T,\mathcal{M}_{n}]$ and theoretical data $[H_{p}/H_{d},\mathcal{M}_{n}]$ we use the phenomenological scaling form of temperature dependence of the critical depinning current density

\begin{equation}
    \label{XTM}
    \frac{J_{c}(T)}{J_{c}(0)} = \left[ 1 - \left( \frac{T}{T_{c}} \right) \right]^{b},
\end{equation}

\noindent with the exponent $b$, typically ranging from 1 to 3 in experiments. By analogy with $J_{c}(T)$ we can define effective temperature $(T/T_{c})_{e} \equiv 1-(cH_{d}/H_{p})^{1/b}$ for theoretical data, where $c \equiv H_{p}/H_{d}(0)=2H_{p}/J_{c}(0)d$ \cite{Janu14}. A relation for effective temperature maps $H_{p}/H_{d}$ from the interval $[H_{p}/H_{d}(0),\infty]$ to the reduced temperature $T/T_{c}$ from the interval $[0,1]$ and vice versa using free parameters $c$, $T_{c}$, and $b$ that fit experimental data to theoretical data and give $J_{c}(0)$ and $J_{c}(T)$. A good agreement including distinctive behavior of the third harmonic is obtained with $T_{c}=$ 8.925 K, $J_{c}(0)=112$ GA/m$^2$, and $b$ = 1.33. The third harmonic components indicate the critical state with $J_{c} > 0$ clearly. Starting by low temperatures and fixed $H_{p}$, with increasing temperature, i.e. increasing $H_{p}/H_{d}$, energy dissipated per ac field cycle increases as $\mathcal{M}''_{1}/m_{M} \propto (H_{p}/H_{d})^2$.
The peak in $\mathcal{M}''_{1}/m_{M}$ has a value consistent with an expected value $\approx 0.24$ and occurs at temperature $T/T_{c} \approx 0.988$ when $H_{p}/H_{d}=1.943$. Further increase in temperature causes decrease in $\mathcal{M}''_{1}/m_{M}$ because the magnetization loop saturates. For $H_{p}/H_{d} \gg 1$ we have $\mathcal{M}''_{1}/m_{M} \propto (H_{p}/H_{d})^{-1}$. Since $J_{c}$ approaches zero as $T$ approaches the critical depinning temperature $T_{c}$, the rate independent $\mathcal{M}''_{1}/m_{M} \propto J_{c}$ turns to the rate dependent $\mathcal{M}''_{1}/m_{M} \propto f \sigma$ with temperature dependence given by temperature dependence of conductivity in the normal state. The critical state is established by presence of the third harmonic components $\mathcal{M}_{3}$. Since a peak in experimental $\mathcal{M}''_{3}/m_{M}$ is not pronounced at $T/T_{c} \approx 0.995$ while the experimental $\mathcal{M}'_{3}/m_{M}$ fits to the predicted curve a clear conclusion cannot be drawn.

Let the applied field synthesize the pure sine wave and its 3rd harmonic, $H(t)=H_{1} \sin (2 \pi f_{1} t) + H_{2} \sin (2 \pi 3f_{1} t)$. For $H_{2}/H_{1} \le 1/9$ the field $H(t)$ increases and decreases monotonously between $-H_{p}$ and $H_{p}$. The added third harmonic trims $H_{p}$ so, in the critical-state, the energy dissipated per cycle of the fundamental field $H_{1}$ decreases with increasing $H_{2}/H_{1}$. At the same time the effective value of the field increases, $H_{rms}=2^{-1/2} \sqrt {H_{1}^{2}+H_{2}^{2}} > 2^{-1/2} H_{1}$. Because of the nonlinear $m(H)$ dependence in the critical-state both $\mathcal{M}(f_{1})$ and $\mathcal{M}(3f_{1})$ depend on both applied fields $H_{1}$ and $H_{2}$ from which energy is absorbed. The total value of the energy dissipation rate is $P = f_{1}W(f_{1})+3f_{1}W(3f_{1})$.

\begin{figure}
\includegraphics[scale=0.5]{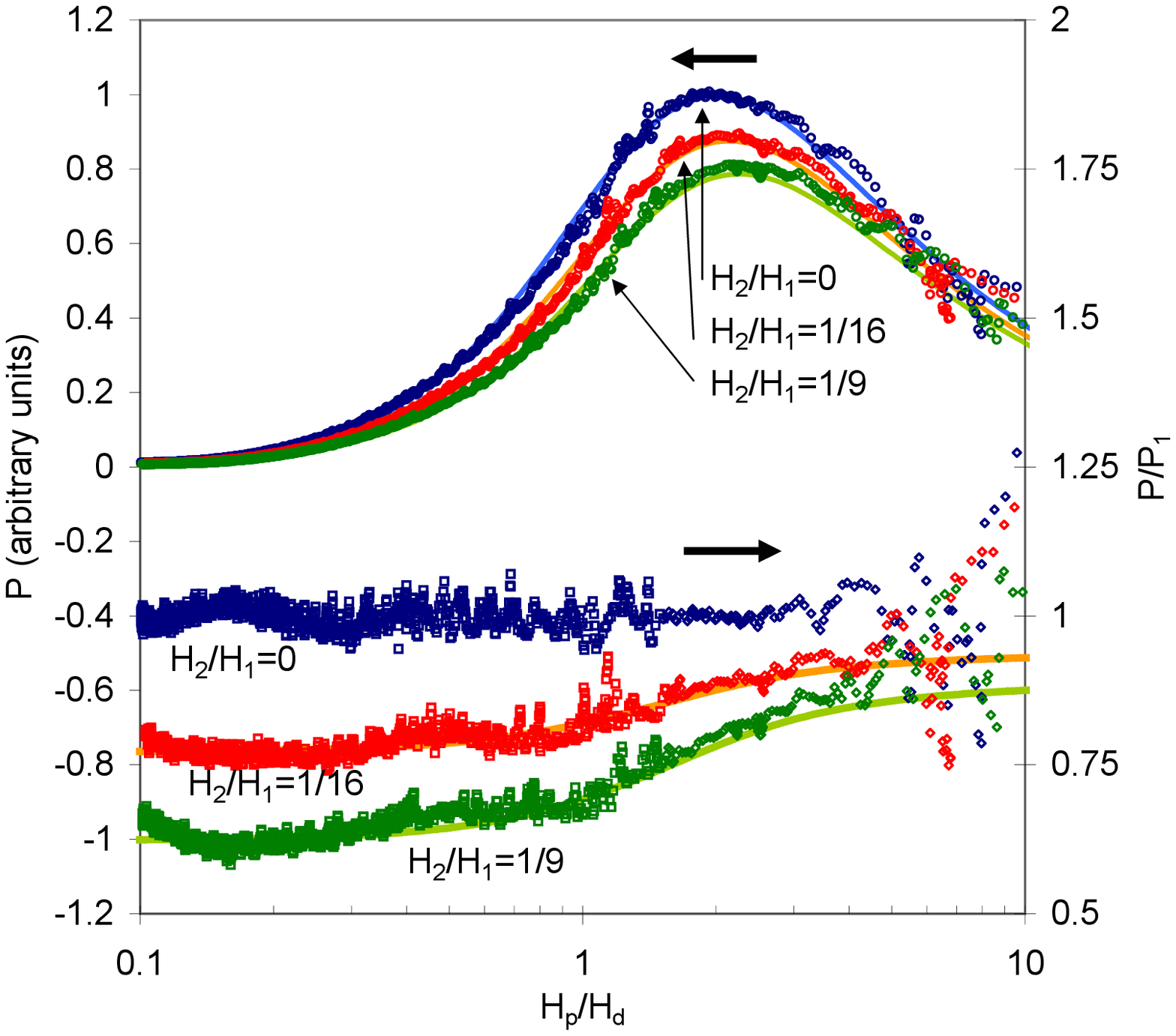}
\caption{\label{FigTDP(HpDivHd)} Top: total value of the energy dissipation rate vs. $H_{p}/H_{d}$ 
in field $H(t) = H_{1}\sin(\omega t) + H_{2}\sin(3\omega t)$ for $\mu_{0}H_{1}=$ 100 $\mu$T and $H_{2}/H_{1} = $ 0, 1/16, and 1/9. Bottom: total value of the energy dissipation rate normalized to the energy dissipation rate in the pure sinusoidal field ($H_{2}=0$). Symbols represent experimental data and curves represent data calculated on the basis of model.}
\end{figure}

Figure \ref{FigTDP(HpDivHd)} shows an impact of the added 3rd harmonic component on the total value of the energy dissipation rate $P$ and normalized $P/P_{1}$ plotted vs. $H_{p}/H_{d}$ for the fixed values $H_{2}/H_{1}=$ 0, 1/16, and 1/9. Here, $P_{1}$ is the energy dissipation rate in the pure sine field $H_{1}$. The experimental $P$ is calculated from the magnetization loops measured at $\mu_{0}H_{1}=$ 100 $\mu$T, $f_{1}=1.5625$ Hz, and varying temperature with rate 0.1 K/min. Temperature dependence is transformed into $H_{p}/H_{d}$ dependence on the basis of $J_{c}(T)$ found using the data shown in figure \ref{FigX1}. Since the field increases and decreases monotonously between $-H_{p}$ and $H_{p}$ we can apply the model for disks. For $H_{p}/H_{d} \ll 1$ the simulation predicts decrease in the energy dissipation rate to $P/P_{1} =$ 0.733 and 0.625 for $H_{2}/H_{1}=$ 1/16 and 1/9, respectively. While the peak in the field is trimmed to $H_{p}/H_{1}=$ 0.9375 and 0.8889 the effective field $H_{rms}$ is increased from 0.707 to 0.708 and 0.711, respectively. An increase in $H_{rms}$ by $\approx 1$\% causes decrease in the energy dissipation rate $P$ by $\approx 40$\%. For $H_{p} \gg H_{d}$ the effect is smaller, $P/P_{1}=$ 0.936 and 0.887 for $H_{2}/H_{1}=$ 1/16 and 1/9, respectively. Figure \ref{FigTDP(HpDivHd)} shows that experimental data are in a good agreement with theoretical predictions including a step in the normalized energy dissipation rate at $H_{p}/H_{d} \approx 2$.

Apparently, the optimum applied field waveform to minimize the total energy dissipation rate in the critical-state and maximize the effective field at the same time is the square waveform as the crest factor $H_{p}/H_{rms}$ of which is the lowest of all waveforms. However, because of experimental limitations e.g. a detection system slew rate and solenoid charging rate we rather synthesize the square waveform field using a series expansion

\begin{equation}
    \label{EqSquareWave}
    H(t) = H_{1} \sum_{n=1,3,5,...}^{N}\frac{\sin \left( n 2 \pi f_{1} t \right)}{n}.
\end{equation}

\noindent As $N$ approaches infinity both $H_{p}$ and $H_{rms}$ approach $(\pi / 4) H_{1} \approx 0.785 H_{1}$ that represents in comparison with the pure sine field an increase in the effective field by 11\%. The total energy dissipation rate in the critical-state is

\begin{equation}
    \label{EqP}
    P = \sum_{n=1,3,5,...}^{N} n f_{1} W \left( nf_{1} \right).
\end{equation}

\noindent In this case the model for disks is not applicable, since the field is not monotonously non-decreasing and non-increasing between the peak values for small $N$. However, a "true" square wave meets this condition so we can evaluate a difference between an energy dissipation rate in sine and square wave fields which have the same effective value. For $H_{p}/H_{d} \ll 1$, $\mathcal{M}''_{1}/m_{M} \propto (H_{p}/H_{d})^2$ that yields $P/P_{1}=0.5$ in favor of the square wave. On the other hand, for $H_{p}/H_{d} \gg 1$, $\mathcal{M}''_{1}/m_{M} \propto (H_{p}/H_{d})^{-1}$ that yields $P/P_{1}=1.4$ increase in the energy dissipation rate. The peak in $P_{1}$ occurs at $H_{rms}/H_{d}=1.55$ while the peak in $P$ occurs at $H_{rms}/H_{d}=2.14$. Both waveforms bring about the same energy dissipation rate at $H_{rms}/H_{d}=1.7$.

In our case with the square waveform given by equation (\ref{EqSquareWave}) the total value of the energy dissipation rate $P=0.616P_{1}$ is decreased by 38\% while the effective field is increased by 11\%.

On the other hand, in the normal state each added sine wave, independently of its phase, increases the total value of the energy dissipation rate. Since the energy dissipation rate increases as a square of the frequency while the amplitudes decrease as a reciprocal value of the frequency,

\begin{equation}
    \label{E5}
    P \propto \sum_{n=1,3,5,...}^N \left( n f_{1} \right) ^{2} \sigma \left(\frac {H_{1}}{n}\right)^{2} \propto \frac{N+1}{2},
\end{equation}

\noindent the total value of the energy dissipation rate increases with number of terms in series expansion if the conductivity is frequency independent.

\begin{figure}
\includegraphics[scale=0.5]{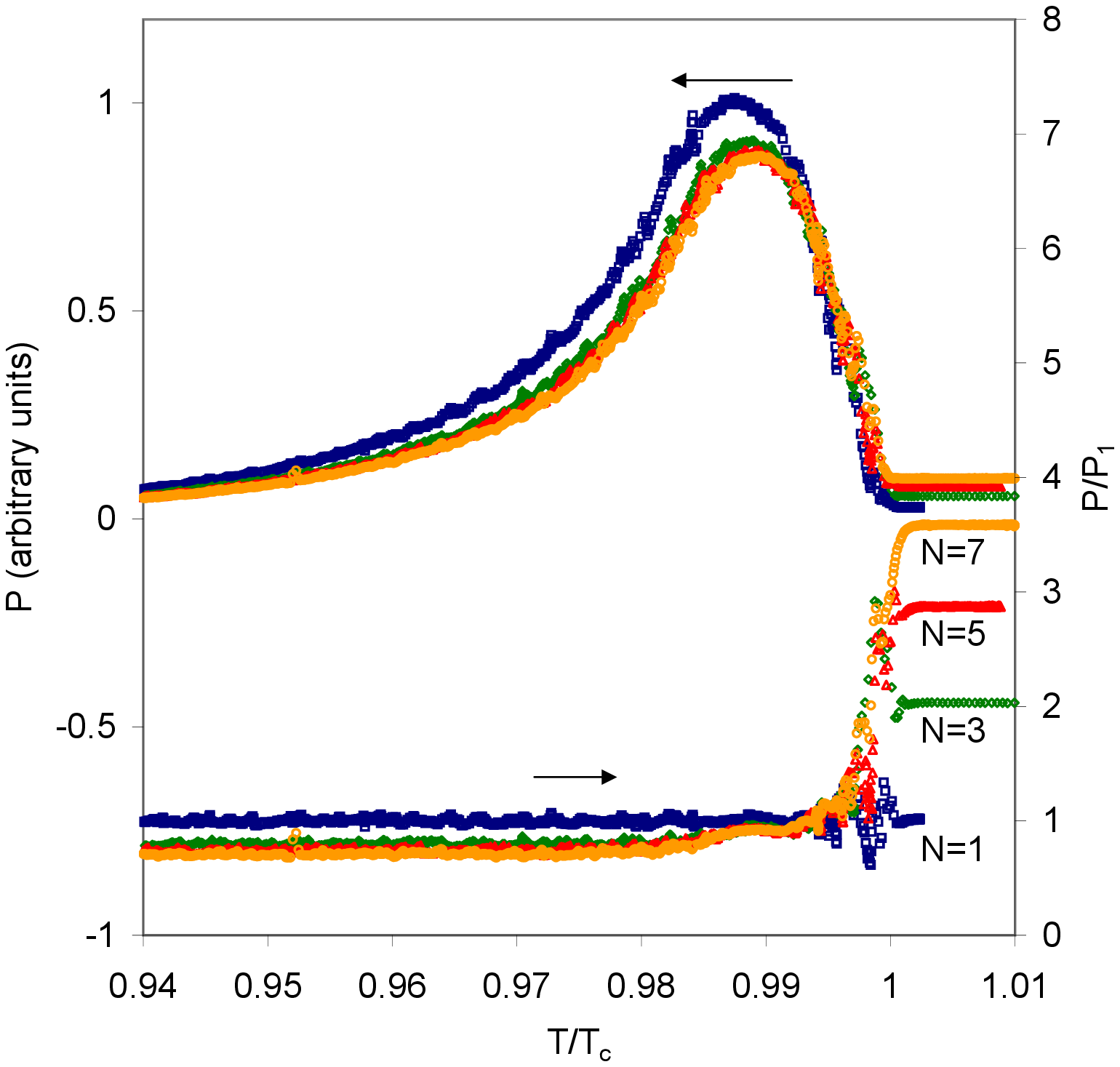}
\caption{\label{FigTDP(TDivTc)} Top: Temperature dependence of the total dissipated 
power for the square wave field synthesized from first 1, 2, 3, and 4 terms in series expansion with $\mu_{0}H_{1}=$ 100 $\mu$T. Bottom: the total energy dissipation rate normalized to the energy dissipation rate in the pure sinusoidal field.}
\end{figure}

Figure \ref{FigTDP(TDivTc)} shows temperature dependencies of the total energy dissipation rate $P$ and normalized total energy dissipation rate $P/P_{1}$ for $N=1$, 3, 5, and 7 measured at $\mu_{0}H_{1}=100$ $\mu$T and $f_{1}=1.5625$ Hz. While the normalized total energy dissipation rate in the critical-state changes only slightly, $P/P_{1}=$ 0.944, 0.942, and 0.941, for $N = 3$, 5, and 7, it increases to $P/P_{1}=$ 2.03, 2.88, and 3.58 in normal state. These values are close to those predicted by the equation (\ref{E5}).

\section{\label{sec:Conclusions}Conclusions}

In conclusion, we have proved experimentally a theoretical prediction that the energy dissipation rate in a quasistatic system with diverging relaxation times excited by a square waveform field is lower than the energy dissipation rate in a pure sine field of the same effective value unlike a system with diffusion dynamics where reverse is true. An analogous affect is observed, for example, when a proper 3rd harmonic of the sinusoidal field that trims the peak value of the field is superposed.

\section*{Acknowledgements}

The magnetization measurements were carried out at the Magnetism and Low-Temperature Laboratories supported within the Program of Czech Research Infrastructures (Project No. M2011025).

{}

\end{document}